\documentclass[12pt]{iopart}

\usepackage{graphicx}

\begin{document}

\title[Modelling Multi-Trait Scale-free Networks]{Modelling Multi-Trait Scale-free Networks by Optimization}
\author{Bojin ZHENG$^{1,2,3*}$, Hongrun WU, Jun QIN$^{1,2}$, Wenfei LAN and Wenhua DU$^{4,*}$}
\address{1. College of Computer Science, South-Central University for Nationalities, Wuhan 430074, China\\
2. State Key Laboratory of  Networking and Switching Technology, Beijing University of Posts and  Telecommunications, Beijing 100876, China\\
3. School of Software, Tsinghua University, Beijing 100084,China\\
4. Management School, South-Central University for Nationalities, Wuhan 430074, China\\
}
\ead{zhengbojin@gmail.com}

\begin{abstract}
Recently, one paper in \emph{Nature}(Papadopoulos, 2012) raised an old debate on the origin of the scale-free property of complex networks, which focuses on whether the scale-free property origins from the optimization or not. Because the real-world complex networks often have multiple traits, any explanation on the scale-free property of complex networks should be capable of explaining the other traits as well. This paper proposed a framework which can model multi-trait scale-free networks based on optimization, and used three examples to demonstrate its effectiveness. The results suggested that the optimization is a more generalized explanation because it can not only explain the origin of the scale-free property, but also the origin of the other traits in a uniform way. This paper provides a universal method to get ideal networks for the researches such as epidemic spreading and synchronization on complex networks.
\end{abstract}

\pacs{89.75.Da, 89.75.Fb, 89.20.Hh}
\noindent{\it Keywords}: Complex Network; Optimization; Multi-trait; Origin; Scale-free Network

\submitto{Arxiv}
\maketitle

\section{Introduction}

Some complex networks present the scale-free property\cite{1}, that is, their degree distribution $p(k)$
satisfies $p(k) \sim k^{-\gamma}$. This phenomenon have been discovered in many complex networks,
such as the Internet\cite{1,221,222}, the World Wide Web\cite{1} and the Scientific Cooperation Network\cite{80}. Owing to the
abroad interests, the researches on complex network involves more and more disciplines\cite{34}.

Up to date, hundreds of models have been proposed to explain the origin and the mechanisms of the scale-free networks.
Among them, the most popular model is the preferential attachment with the growth (the BA model)\cite{1}. This model says
that all scale-free networks are evolving; when new nodes emerge, they will link to the previous nodes
with a linear preference which is proportional to the degrees of the previous nodes. 

Notice that the real-world networks commonly are multi-trait. That is, one network can be
not only scale-free, but also small-world\cite{16} or has other traits\cite{116,145}. Therefore, every explanation
on the origin of the scale-free networks should be capable of explaining the other traits, otherwise,
it can not be treated as the ultima explanation.
Because the preferential attachment with the growth only explains the mechanism of the scale-free property,
and it can not explain the other traits such as the small-world effect and fractal structure\cite{145}, researchers
developed many explanations based on the preferential attachment.

Actually, in spite of the popularity of the preferential attachment, researchers have proposed
other explanations, such as the randomness\cite{94}, the optimization\cite{37,38}, the hierarchy\cite{47,151}. For example, Carlson et al. thought
that the preferential attachment can not explain the Internet and proposed that the HOT (Highly Optimized Tolerance) is the right mechanism\cite{37,38}.

Recently, an old debate arose again\cite{224}. On one side, Papadopoulos et al. proposed the popularity $\times$ similarity
optimization model to explain the origin of the preferential attachment and the scale-free networks\cite{225};
Zheng et al. proposed a dual optimization model to explain the origins of complex networks and to
clarify their relationships\cite{zheng317}. These results show that the scale-free property and the other traits
can origin from the optimization. On the other side, the BA model still dominates with strong supports from evidences and theories.
Therefore, whether the scale-free networks origins from the the
randomness or the optimization? Furthermore, whether the preferential attachment origins from
the randomness or the optimization? These questions require the answers.


In this paper, we demonstrate that the power law distribution
can be transferred into the optimization problems and then propose an optimization framework.
Under this framework, we can append additional traits such as the small-world effect,
high clustering coefficient et al. into the resultant scale-free networks.
This paper validates the optimization explanation on the origin of the scale-free networks and even most of complex networks with compatibility to the BA model . Moreover, this paper provides a universal method to the generation of the ideal multi-trait networks for the simulation experiments
in the other researches, such as the invulnerability, epidemic spreading, game theory and synchronization on complex networks.

\section{Methods}

In this paper, we transfer the power law distribution into the optimization objective, and then add the other traits
as the constraints, therefore, the solutions of the obtained optimization problem would be scale-free networks with
the specific traits. When the optimization problem is determined, we can develop an algorithm to solve it, i.e., obtain the desired network.

\subsection{The optimization objective}

To generate a scale-free network, there exist many methods, and these methods often are regarded as the mechanisms to explain how the scale-free networks origin. From the view of optimization, there also are many methods to generate the scale-free property. Here we introduce a simple method, that is, treating the scale-free property as the optimization objective.

When treating the scale-free property as the optimization objective, we actually optimize the resultant network to make its degree distribution to satisfy the power law distribution. Assume that the desired network is undirected and unweighted with $N$ nodes, we actually need to make the degrees of nodes to be the samples of the power law distribution. In another words, we can generate $N$ ideal samples (see the appendix for the method), and then optimize a network to achieve that its node degrees match these samples. That is, the optimization objective can be written as equation \ref{eqnobj}.

\begin{equation}\label{eqnobj}
    min \ g(A) = \sum\limits_{i = 1}^N { {(d_i(A)   - s_i )}^2}
\end{equation}

Here, $A$ is the adjacent matrix of the network, $d_i$ is the degree of the i-th node, and $s_i$ is the value of the i-th sample.

\subsection{The constraints}

The other traits of the desired networks can be treated as the constraints. For example, the clustering coefficient plays an important role in complex networks, if we want the network has a specific clustering coefficient $\mathcal{C}$, then the optimization problem can be expressed as equation \ref{eqnobjcoeff}.
\begin{equation}\label{eqnobjcoeff}
    \begin{array}{l}
    min \ g(A) = \sum\limits_{i = 1}^N { {(d_i(A)   - s_i )}^2}\\
    \mbox{s.t.}\\
    cc(A)= \mathcal{C}
 \end{array}
\end{equation}

Here $cc$ is the clustering coefficient function of the optimizing network.

Moreover, the average shortest path length is a key traits for complex networks, especially, many complex networks present a characteristic value, about $ln(N)$.
 We can express the the average shortest path length to be a constraint as equation \ref{eqn.asp}.

\begin{equation}\label{eqn.asp}
    \begin{array}{l}
    min \ g(A) = \sum\limits_{i = 1}^N { {(d_i(A)   - s_i )}^2}\\
    \mbox{s.t.}\\
    y(A)=ln(N)
 \end{array}
\end{equation}

Here $y$ is the average shortest path function of the optimizing network.

Of course, one optimization problem can have multiple constraints. Taking the small-world effect as the example, the small-world networks should include two constraints, that the average shortest path length is $ln(N)$ and the clustering coefficient is medium. Therefore, it can be depicted as equation \ref{eqn.smallworld}.
 
\begin{equation}\label{eqn.smallworld}
    \begin{array}{l}
    min \ g(A) = \sum\limits_{i = 1}^N { {(d_i(A)   - s_i )}^2}\\
    \mbox{s.t.}\\
    y(A)=ln(N)\\
        cc(A)= \mathcal{C}\\
    
 \end{array}
\end{equation}

\subsection{The optimization algorithm}

Once the form and parameters of the optimization problem are finally determined, we can employ an optimization algorithm to solve this problem.

According to the Lagrangian
relaxation method\cite{nlp}, equation \ref{eqnobjcoeff} can be rewritten as equation \ref{eqn.eqnobjcoeff.2}.

\begin{equation}\label{eqn.eqnobjcoeff.2}
    min \ g'(A) = \sum\limits_{i = 1}^N { {(d_i(A)   - s_i )}^2} + \theta {(cc(A)- \mathcal{C})}^2
\end{equation}

Here $\theta$ is an arbitrary positive real number.

Similarly,
equation \ref{eqn.asp} can be rewritten as equation \ref{eqn.asp.2}.

\begin{equation}\label{eqn.asp.2}
    min \ g'(A) = \sum\limits_{i = 1}^N { {(d_i(A)   - s_i )}^2} + \theta {(y-ln(N))}^2
\end{equation}

When the optimization model has multiple constraints, the Lagrangian
relaxation method needs multiple parameters. So that equation \ref{eqn.smallworld} can be rewritten as equation \ref{eqn.smallworld.2}.

\begin{equation}\label{eqn.smallworld.2}
    min \ g'(A) = \sum\limits_{i = 1}^N { {(d_i(A)   - s_i )}^2} + \theta {(y-ln(N))}^2 + \varphi {(cc(A)- \mathcal{C})}^2
\end{equation}

Because the optimization problems are simple, we employ the classic hill-climbing algorithm to solve them. Because the hill-climbing algorithm belongs to the iterative algorithm, this paper set the iteration number as 100000.

The algorithm can be depicted as Fig. \ref{fig.code}.

\begin{figure}[htbp]
\begin{flushleft}
0.Generating the samples;\\
1.Initializing a connected network $A$ randomly and let count=0;\\
2.Calculating $g'(A)$;\\
3.while count $<$ 100000\\
4.\ \ \ $A'$ = $A$ and change an edge of $A'$ randomly;\\
5.\ \ \ if $A'$ is disconnected then jump to 3;\\
6.\ \ \ Calculating $g'(A')$;\\
7.\ \ \ if $g'(A')$ $<$ $g'(A)$\\
8.\ \ \ \ \ \ A = A';\\
9.\ \ \ end if;\\
a.\ \ \ count = count + 1;\\
b.end while;\\
c.Output $A$.\\
\end{flushleft} \caption{The pseudocode of the optimization algorithm}\label{fig.code}
\end{figure}

\section{Results}

We solved equation \ref{eqnobjcoeff}, \ref{eqn.asp} and \ref{eqn.smallworld} with different parameters by simulation. We select 9 groups of parameter settings that are listed as Table \ref{tab.para}. Moreover, all the number of nodes are set to 300 such that the topology of the resultant networks can be clearly visualized.
\begin{table}[h]
  \centering
  \caption{The parameter settings}
  \(\begin{array} {|p{30pt}|p{12pt}|p{24pt}|p{24pt}|p{24pt}|p{24pt}|p{24pt}|p{24pt}|p{24pt}|}
  \hline
  Group& $\gamma$ & $kmin$ &$kmax$ &$E$& $\theta$& $\varphi$ & $l$ & $\mathcal{C}$\\
\hline
(A)& 	2&	 1&	27&	347&	1&	-&	-	&0.06\\
(B)&	2&	 1&	27&	347&	1&	-&	-   &0.1\\
(C)&    2&   2&	43&	761&	1&	-&	-   &0.6\\
(D)&	2&	 1&	27&	347&	-&	1&	5.7&-\\
(E)&	2&	 2&	43&	347&	-&	1&	5.7&-\\
(F)&	2.4& 2&	30&	559&	-&	1&  5.7&-\\
(G)&	2&	 1&	27&	347&	1&  1&  5.7&0.1\\
(H)&	2&	 2&	43&	761&	1&  1&  5.7	&0.6\\
(I)&	2.4& 2&	30&	559&	1&  1&  5.7	&0.3\\
\hline
\end{array}\)
  \label{tab.para}
\end{table}

In Table \ref{tab.para}, $l$ is the expected average shortest path, $E$ is the number of the edges of the resultant network, $\gamma$ is the expected exponent of the power law distribution, $kmax$ is the maximal value of node degrees.

In the experiments, we carried out 30 times for every parameter setting. The corresponding statistical results are listed as Table \ref{tab.rst}.
 
\begin{table}[h]
  \centering
  \caption{The corresponding results}
  \(\begin{array} {|p{30pt}|p{42pt}|p{42pt}|p{42pt}|p{48pt}|p{42pt}|p{48pt}|}
  \hline
  Group& $Avg(\gamma')$ & $STD(\gamma')$ &$Avg(y)$ &$STD(y)$& $Avg(cc)$& $STD(cc)$\\
\hline
(A)&	2.05692&	0.00751&	-&	-&	0.06000&	3.60E-07\\
(B)&	2.08833&	0.00389&	-&	-&	0.10001&	7.78E-07\\
(C)&	2.15800&	0.00389&	-&	-&	0.59510&	0.00969\\
(D)&	2.04800&	0.00315&	5.69990&	1.29E-05&	-&	-\\
(E)&	2.10400&	0.01888&	5.70000&	0&	-&	-\\
(F)&	2.46100&	0.00447&	5.70000&	0&	-&	-\\
(G)&	2.10500&	0.00548&	5.68160& 	0.04635&	0.09535&	0.00603\\
(H)&	2.15000&	0.00408&	5.69439&	0.00408&	0.59442&	0.00283\\
(I)&	2.53800&	0.00447&	5.70001&	3.13E-05&	0.29822&	0.00371\\
\hline
\end{array}\)
  \label{tab.rst}
\end{table}

In Table \ref{tab.rst}, $\gamma'$ is the exponent of the power law distribution of the resultant network, Avg means the average value and STD means the standard deviation.

From Table \ref{tab.rst}, we can see that the experimental results are satisfactory. The indicators of the traits of the resultant networks approximate to the desired value. 

\begin{figure}[h]
  \centering
  \includegraphics[width=8cm, height=6cm]{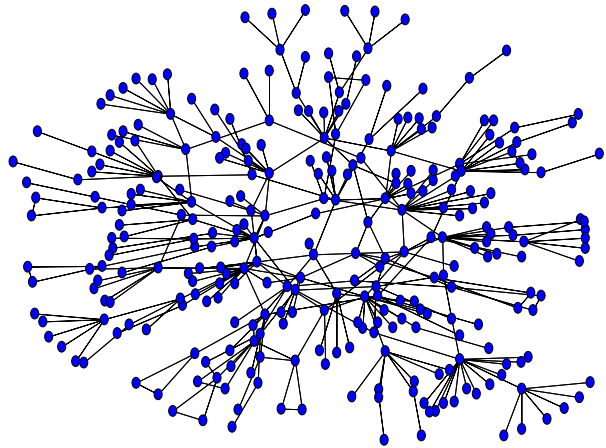}\\
  \caption{The topology of the selected network from group (A)}\label{fig.topology.1}
\end{figure}

\begin{figure}[h]
  \centering
  \includegraphics[width=8cm, height=6cm]{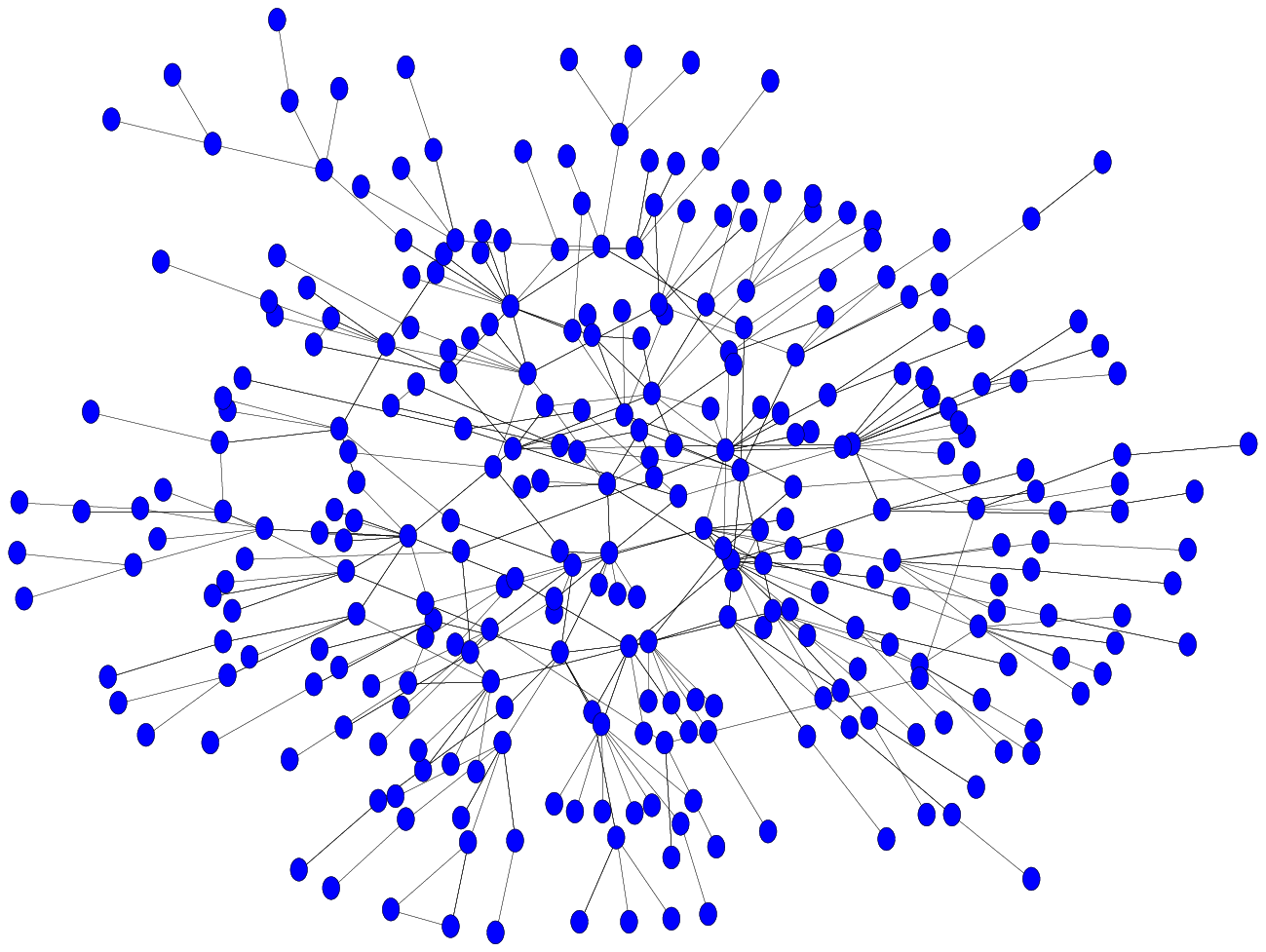}\\
  \caption{The topology of the selected network from group (B)}\label{fig.topology.2}
\end{figure}

\begin{figure}[h]
  \centering
  \includegraphics[width=8cm, height=6cm]{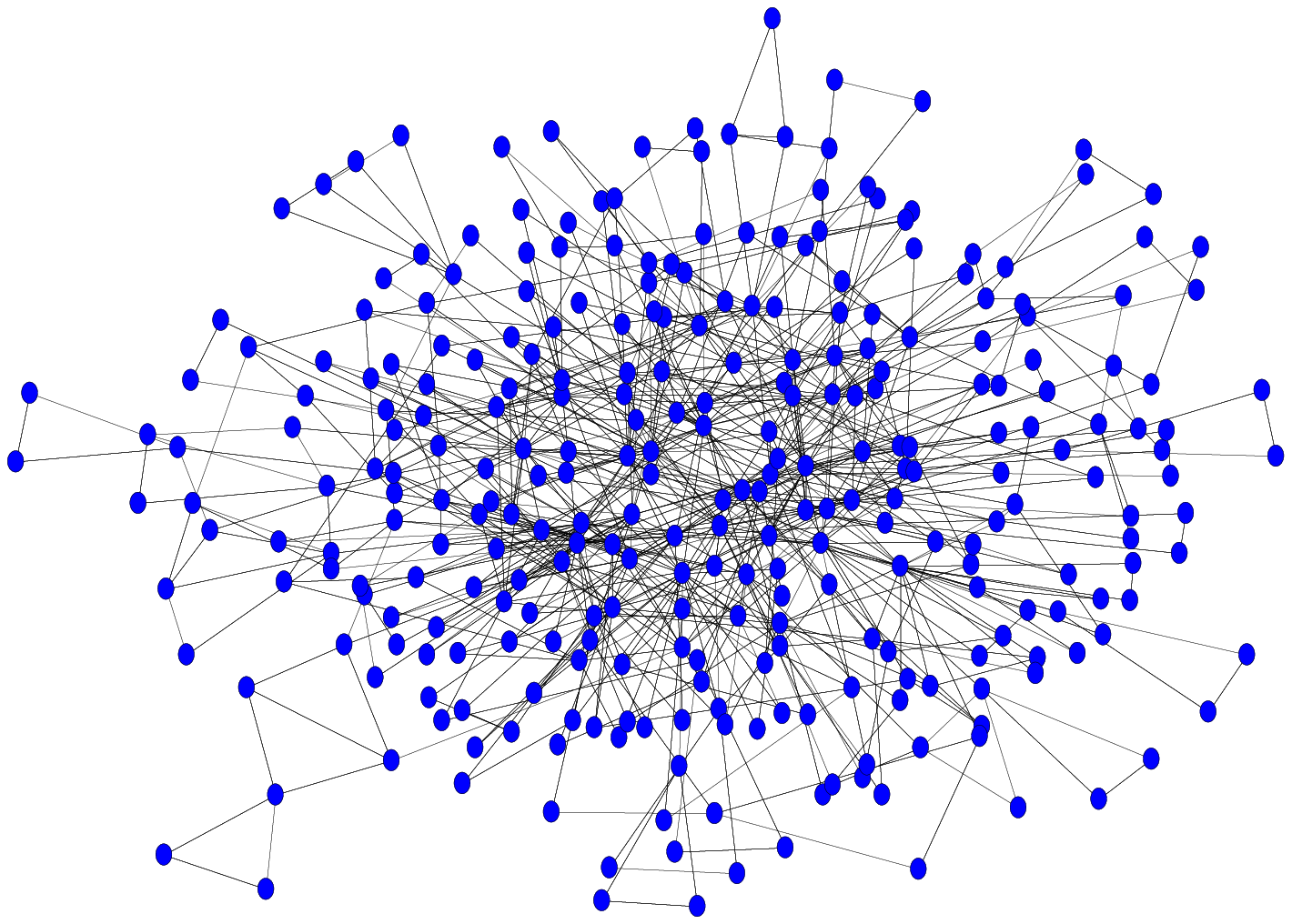}\\
  \caption{The topology of the selected network from group (C)}\label{fig.topology.3}
\end{figure}

\begin{figure}[h]
  \centering
  \includegraphics[width=8cm, height=6cm]{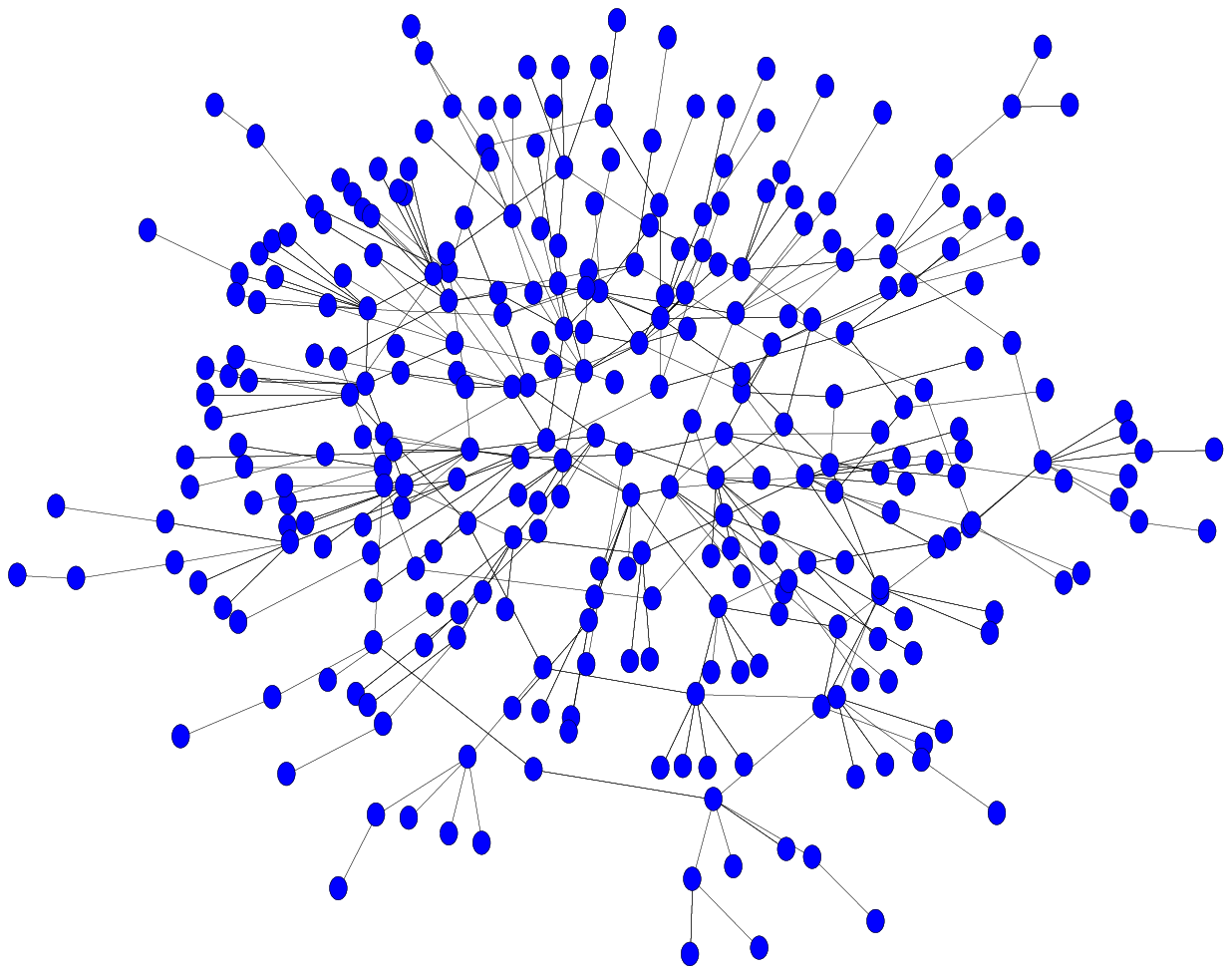}\\
  \caption{The topology of the selected network from group (D)}\label{fig.topology.4}
\end{figure}

\begin{figure}[h]
  \centering
  \includegraphics[width=8cm, height=6cm]{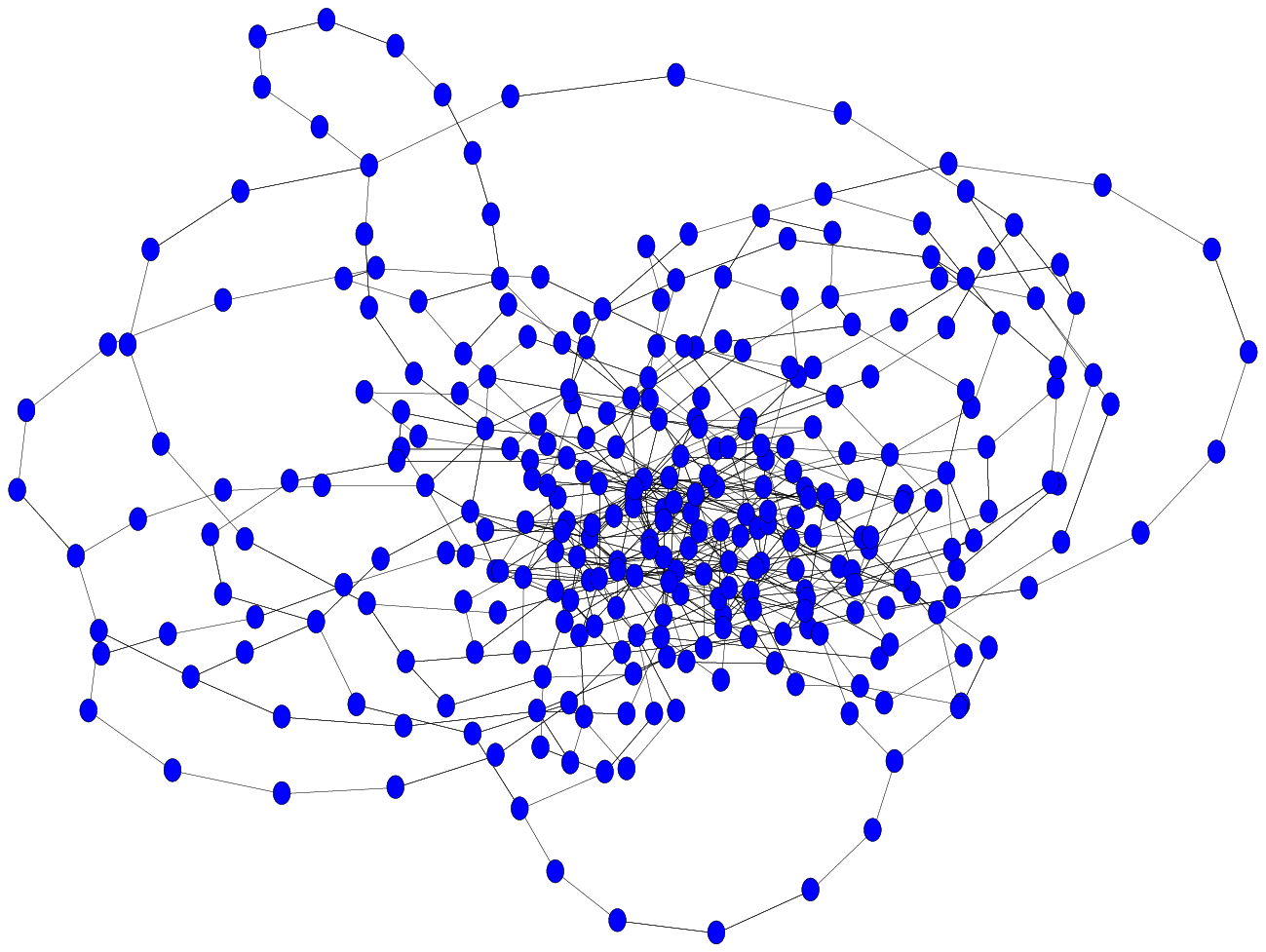}\\
  \caption{The topology of the selected network from group (E)}\label{fig.topology.5}
\end{figure}

\begin{figure}[h]
  \centering
  \includegraphics[width=8cm, height=6cm]{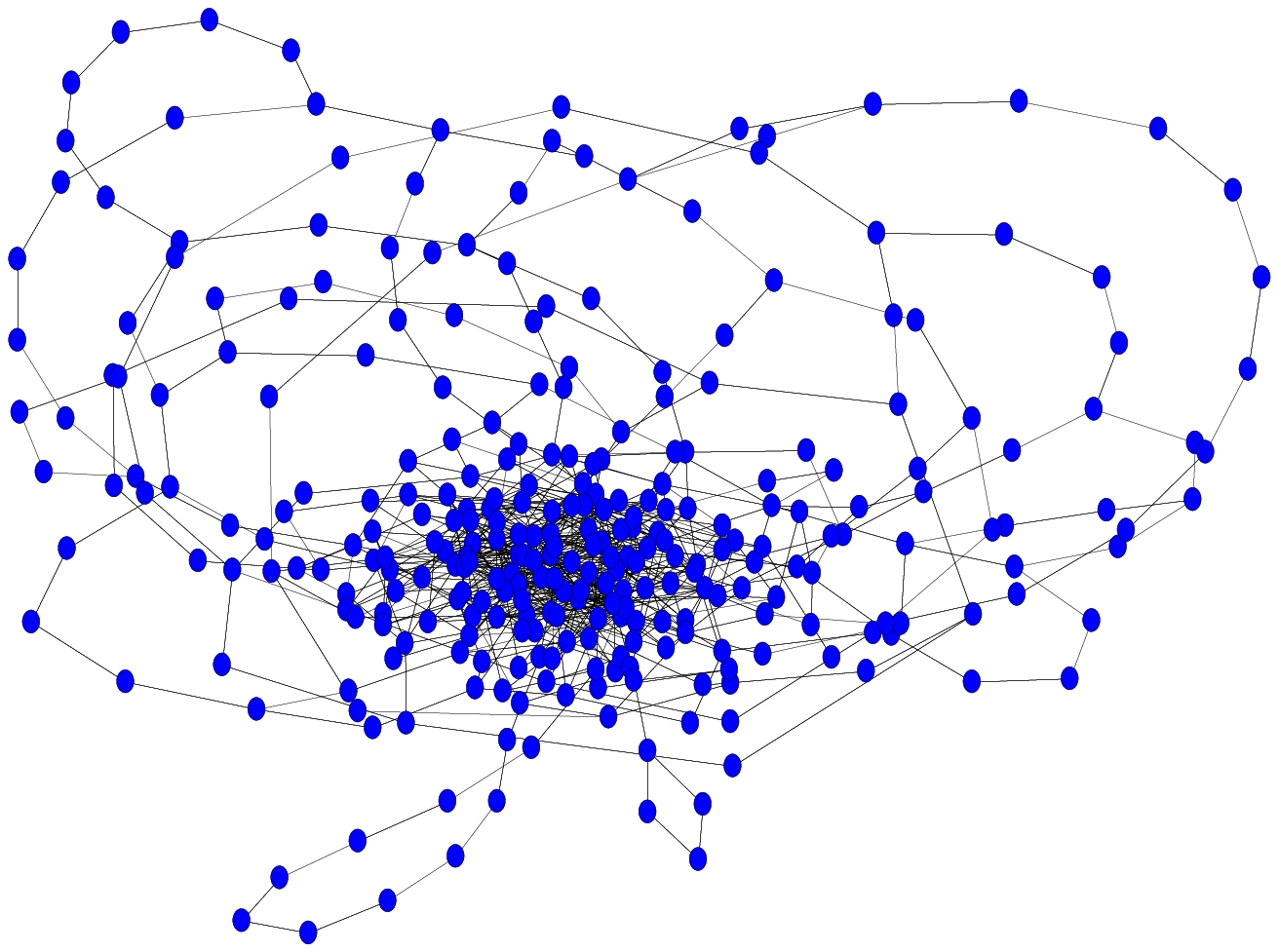}\\
  \caption{The topology of the selected network from group (F)}\label{fig.topology.6}
\end{figure}

\begin{figure}[h]
  \centering
  \includegraphics[width=8cm, height=6cm]{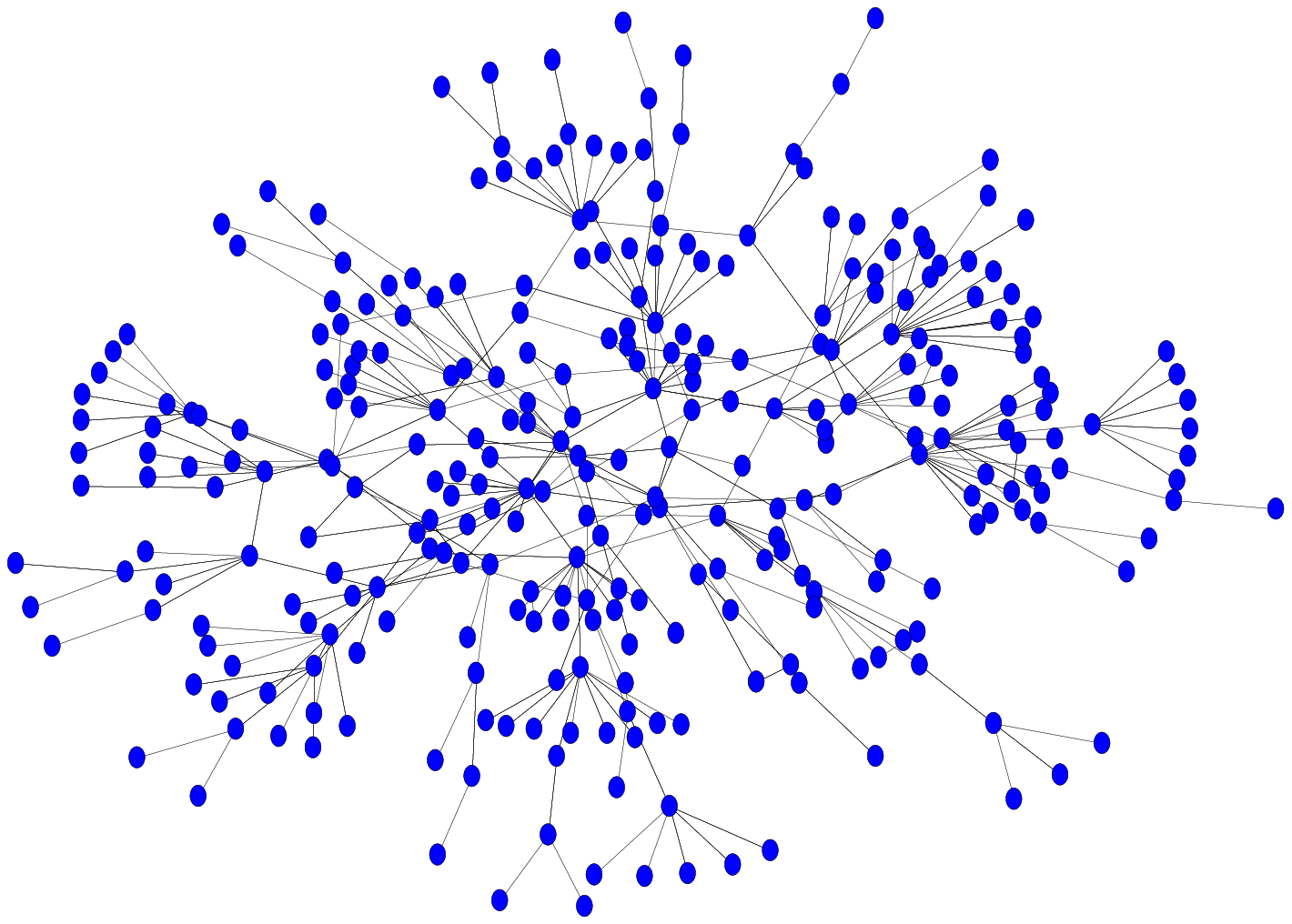}\\
  \caption{The topology of the selected network from group (G)}\label{fig.topology.7}
\end{figure}

\begin{figure}[h]
  \centering
  \includegraphics[width=8cm, height=6cm]{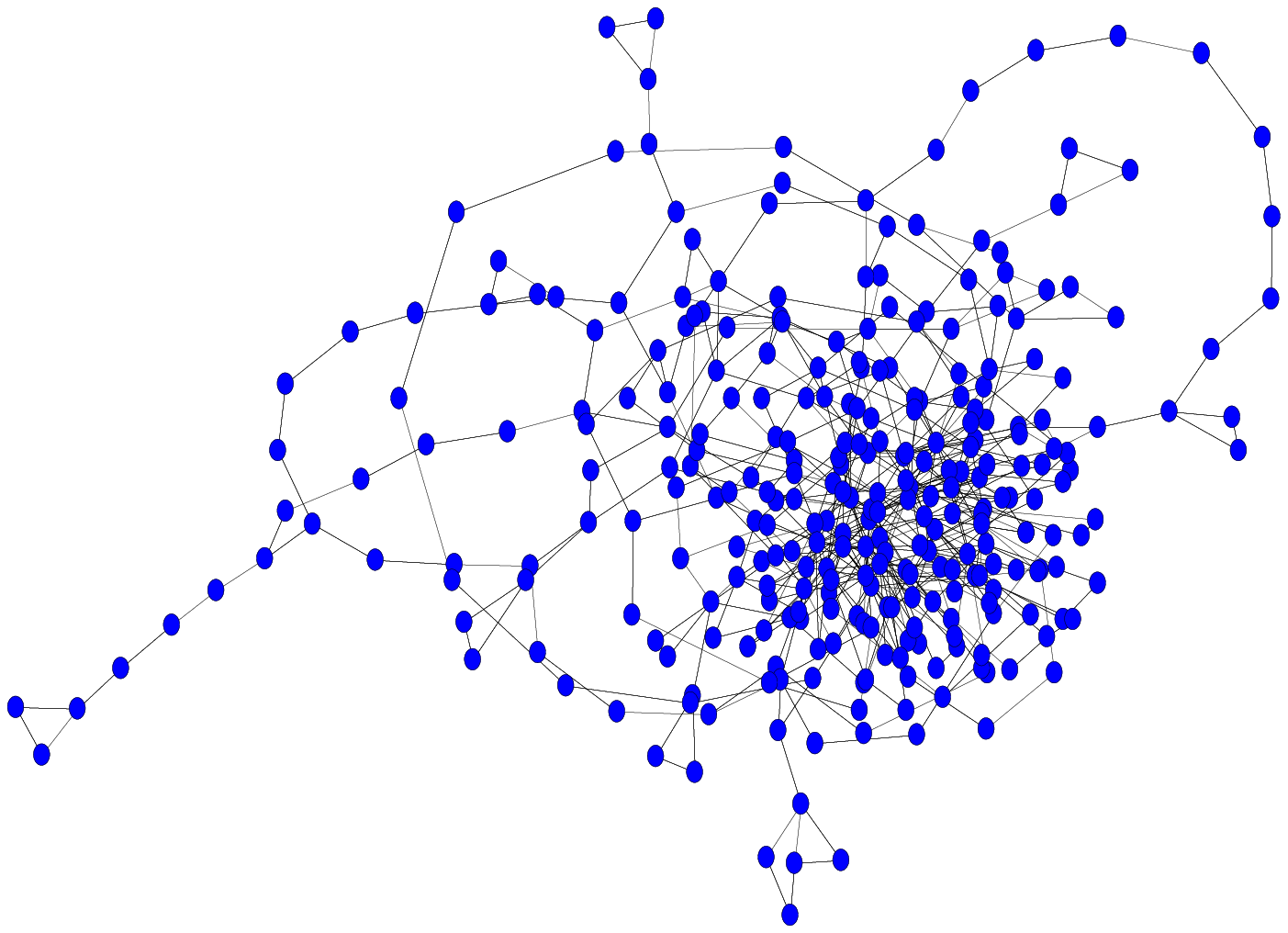}\\
  \caption{The topology of the selected network from group (H)}\label{fig.topology.8}
\end{figure}

\begin{figure}[h]
  \centering
  \includegraphics[width=8cm, height=6cm]{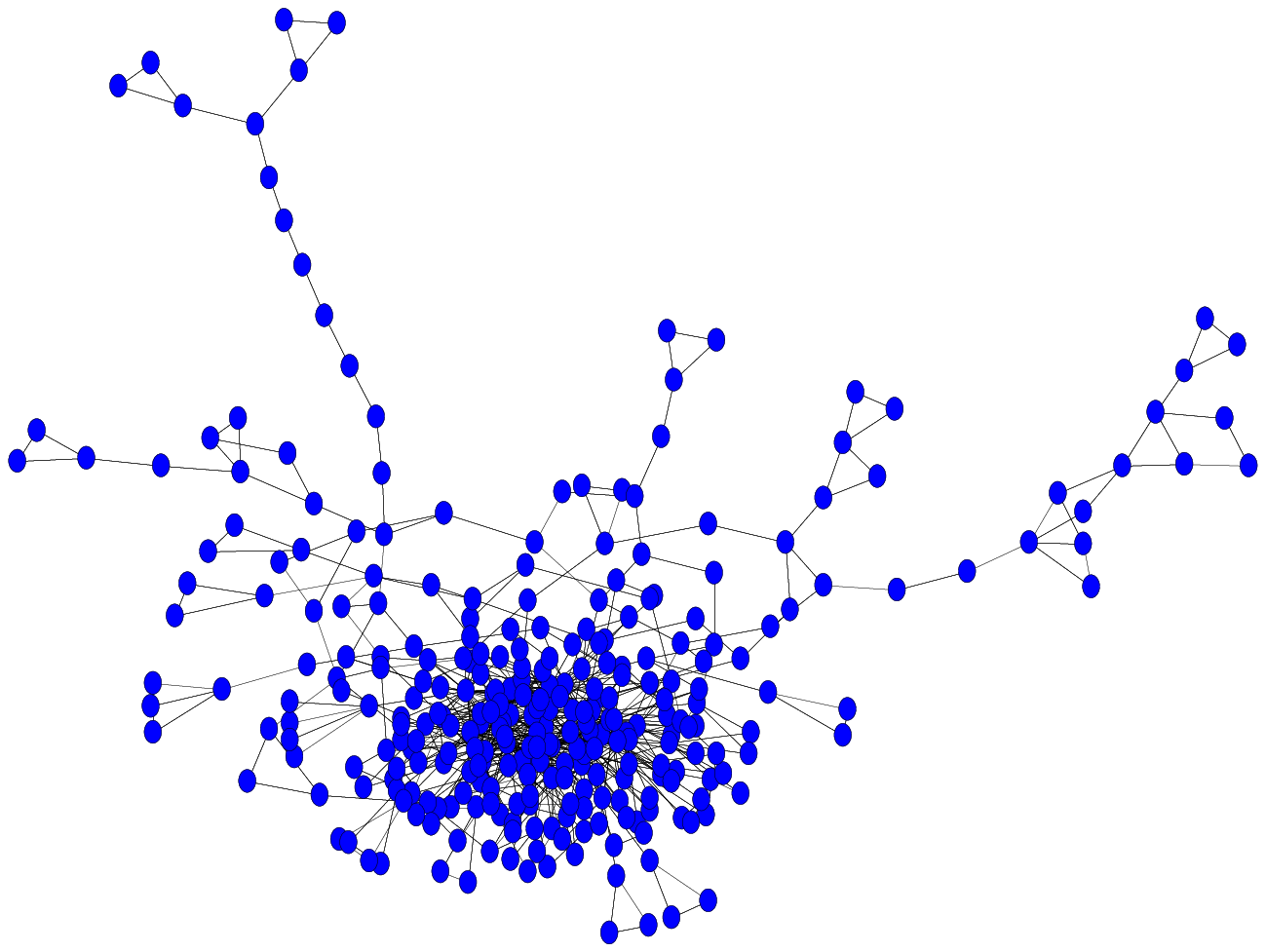}\\
  \caption{The topology of the selected network from group (I)}\label{fig.topology.9}
\end{figure}

For each parameter setting, we chose one resultant network shown as Fig. \ref{fig.topology.1} to \ref{fig.topology.9}.

Comparing Fig. \ref{fig.topology.1} and  \ref{fig.topology.2}, we can see that when the clustering coefficient increases, the edges tend to congregate to the center.

Comparing Fig. \ref{fig.topology.1} to  \ref{fig.topology.3}, we can see that when the $kmin$ increases, the clustering coefficient increases rapidly.

Comparing Fig. \ref{fig.topology.4} and  \ref{fig.topology.5}, we can see that when the $kmin$ increases, the topology of the desired networks change greatly. The edges congregate to the center with long loops around.

Comparing Fig. \ref{fig.topology.5} and  \ref{fig.topology.6}, the center has more dense edges, and the loops are longer.

Comparing Fig. \ref{fig.topology.7} to \ref{fig.topology.9}, we can see that when the $kmin$ and $\gamma$ change, the small-world networks have very different topological structures.

\section{Discussion}

This paper demonstrated the ability of the proposed framework to generate the desired multi-trait networks. According to the optimization theory, the Lagrangian
relaxation method can deal with multiple constraints, therefore, this framework can be easily extend to obtain the networks with more than 2 traits. For example, we can generate a scale-free network with specific average shortest path and clustering coefficient.

Because the proposed framework based on optimization can generate multi-trait networks, we can say that the optimization is an explanation to the origin or the mechanism of the scale-free networks with some traits. this framework only focus on the final forms of the power law distributions, therefore, no matter what are the origin and the mechanisms of the scale-free property, they can be easily integrated in. Furthermore, it has been proved that the scale-free property can be obtained by the optimization, so scale-free networks with arbitrary traits can be expressed as the optimization problems. That is, the optimization is a universal explanation on the origins of scale-free networks.

As to the BA model, it is compatible of the proposed framework, because it can be used only to explain the generation of the scale-free property.

Besides, this paper employed a classic algorithm to solve the optimization problems. If better the-state-of-the-art algorithms are used, the experimental results are expected to be better.

\section{Conclusion}

This paper proposed a general framework to obtain multi-trait scale-free networks. This framework firstly transfers the power law distribution into the optimization objective and the other traits as the constraints, so obtains an optimization problem, and then employs an optimizer to solve this problem, finally obtains the desired multi-trait network. Taking three examples, this paper also demonstrated how to apply this framework to generate the desired network. From the experimental results, we also found that the small-world effect actually means very different topological structures.

Because the proposed framework can uniformly explain the origin of complex networks with multiple traits, this paper provided a perspective on the origin and the mechanisms of complex networks. Especially, this paper can easily integrate the BA model or the optimization explanations into the proposed framework, i.e., this framework is compatible of the BA model and the optimization.

Moreover, this paper can be used to obtain ideal topology of multi-trait complex networks which would be necessary in the researches on the invulnerability of the complex networks\cite{27, 117,152}, the synchronization, the control, the game and the epidemic spreading on complex networks\cite{34,33}.

\section{Appendix}

When $N$ samples are generated, they are expected to satisfy the power law distribution as possible as they can, especially when $N$ is not large enough. Therefore, this paper suggests a method to generate the samples.

Notice the continuous version of the power law distribution, shown as equation (\ref{eq9}), the power law distribution depends on two parameters, the $kmin$ and $\gamma$. Here $kmin$ is the minimum value of the node degrees.

\begin{equation}
\label{eq9}
p(k) = \frac{\gamma - 1}{kmin}\left( {\frac{k}{kmin}} \right)^{ - \gamma }
\end{equation}

When $kmin$ and  $\gamma$ are determined, we can calculate the expected occurrence of every $k$. Although $k$ can be any value between $N - 1$ and $kmin$, in most circumstance, the probability is quite small when $k$ is large, therefore, this paper suggest a maximum value $kmax$ and set the probability of $k$ as 0 when $k$ is larger than $kmax$.  Hence, the power law distribution can be written as equation \ref{eq12}.

\begin{equation}
\label{eq12}
p(k) = \frac{1}{\sum\limits_{k = kmin}^{kmax} {k^{ - \gamma}} }k^{ - \gamma}
\end{equation}

According to equation \ref{eq12}, we should determine $kmax$. By calculating the expected occurrence for each $k$, we can get the probability values for the degrees. If the expected occurrence of certain degree is smaller than a threshold 0.3, then we truncate it, so that we get the $kmax$. Then we recalculate $p(k)$ and round the expected of each $k$ and get the samples.

\ack
JQ is grateful for support from the Fundamental Research Funds for the Central Universities (No. CZY12032) and Nature Sience Foundation in Hubei (No.BZY11010). BZ is grateful for support from the State Key Laboratory of Networking and Switching Technology (No. SKLNST-2010-1-04) and the National Natural Science Foundation of China (No.61273213) and (No.60803095).


\section*{References}
\bibliographystyle{plain}
\bibliography{ComplexNetwork}

\begin{thebibliography}{10}

\bibitem{117}
R\'{e}ka Albert and Albert-L\'{a}szl\'{o} Barab\'{a}si.
\newblock Statistical mechanics of complex networks.
\newblock {\em Reviews Of Modern Physics}, 74(1):47--97, 2002.

\bibitem{224}
A.-L. Barabasi.
\newblock Network science: Luck or reason.
\newblock {\em Nature}, 489(7417):507--508, 2012.

\bibitem{1}
A.~L. Barab\'{a}si and R.~Albert.
\newblock Emergence of scaling in random networks.
\newblock {\em Science}, 286(5439):509--512, 1999.

\bibitem{nlp}
Dimitri~P. Bertsekas.
\newblock {\em Nonlinear Programming: 2nd Edition}.
\newblock Athena Scientific, 1999.

\bibitem{33}
S.~Boccaletti, V.~Latora, Y.~Moreno, M.~Chavez, and D.-U Hwang.
\newblock Complex networks: structure and dynamics.
\newblock {\em physics reports}, 424:175--308, 2006.

\bibitem{37}
J.~M. Carlson and J.~Doyle.
\newblock Highly optimized tolerance: A mechanism for power laws in designed
  systems.
\newblock {\em Phys. Rev. E.}, 60:1412--1427, 1999.

\bibitem{38}
J.~M. Carlson and J.~Doyle.
\newblock Highly optimized tolerance: Robustness and design in complex systems.
\newblock {\em Phys. Rev. Lett.}, 84:2529--2532, 2000.

\bibitem{221}
Michalis Faloutsos, Petros Faloutsos, and Christos Faloutsos.
\newblock On power-law relationships of the {I}nternet topology.
\newblock In {\em Proceedings of the ACM SIGCOMM}, pages 251--262, Cambridge,
  MA, USA, 1999. ACM.

\bibitem{116}
M.~Girvan and M.~E.~J. Newman.
\newblock Community structure in social and biological networks.
\newblock {\em Proc. Natl. Acad. Sci. USA}, 99:8271--8276, 2002.

\bibitem{94}
Wentian Li.
\newblock Random texts exhibit zipf's-law-like word frequency distribution.
\newblock {\em IEEE Trans. Inf. Theory}, 38(6):1842--1845, 1992.

\bibitem{34}
M.~E.~J. Newman.
\newblock the structure and function of complex networks.
\newblock {\em SIAM Review}, 45:167--256, 2003.

\bibitem{27}
M.~E.~J. Newman.
\newblock Power laws, pareto distributions and zipf's law.
\newblock {\em CONTEMP PHYS}, 46:323--351, 2005.

\bibitem{80}
M.E.J. Newman.
\newblock The structure of scientific collaboration networks.
\newblock {\em Proceedings of the National Academy of Sciences},
  98(2):404--409, 2001.

\bibitem{225}
Fragkiskos Papadopoulos, Maksim Kitsak, M.~Angeles Serrano, Marian Boguna, and
  Dmitri Krioukov.
\newblock Popularity versus similarity in growing networks.
\newblock {\em Nature}, 489(7417):537--540, 2012.

\bibitem{47}
Erzs\'{e}bet Ravasz and A.-L. Barab\'{a}si.
\newblock Hierarchical organization in complex networks.
\newblock {\em Phys. Rev. E}, 67:026112, 2003.

\bibitem{145}
Chaoming Song, Shlomo Havlin, and Hernan~A. Makse.
\newblock Origins of fractality in the growth of complex networks.
\newblock {\em Nature Physics}, 2:275--281, 2006.

\bibitem{16}
Duncan~J. Watts and Steven~H. Strogatz.
\newblock Collective dynamics of `small-world' networks.
\newblock {\em Nature}, 393(6684):440--442, 1998.

\bibitem{152}
Bojin Zheng, Dan Huang, Deyi Li, Guisheng Chen, and Wenfei Lan.
\newblock Some scale-free networks could be robust under the selective node
  attacks.
\newblock {\em Europhysics Letters}, 94:28010, 2011.

\bibitem{151}
Bojin Zheng, Jianmin Wang, Guisheng Chen, Jian Jiang, and Xianjun Shen.
\newblock Hidden tree structure is a key to the emergence of scaling in the
  world wide web.
\newblock {\em Chin. Phys. Lett.}, 28(1):018901, 2011.

\bibitem{zheng317}
Bojin Zheng, Hongrun Wu, Jun Qin, Wenhua Du, Jianmin Wang, and Deyi Li.
\newblock A simple model clarifies the complicated relationships of complex
  networks, 2012.
\newblock arXiv:1210.3121.

\bibitem{222}
S.~Zhou and R.J. Mondragon.
\newblock The rich-club phenomenon in the {I}nternet topology.
\newblock {\em IEEE Communications Letters}, 8(3):180--182, 2004.

\end{thebibliography}

\end{document}